\documentclass[conference]{IEEEtran}
\IEEEoverridecommandlockouts
\usepackage{cite}
\usepackage{amsmath,amssymb,amsfonts}
\usepackage{algorithmic}
\usepackage{graphicx}
\usepackage{textcomp}
\usepackage{xcolor}
\usepackage{kotex}
\usepackage{verbatim}
\usepackage{float}
\usepackage[flushleft]{threeparttable} % Table foot note
\usepackage{epsfig}
\usepackage{booktabs}
\usepackage{multirow}

\usepackage[caption=false,font=normalsize,labelfont=sf,textfont=sf]{subfig}

\def\BibTeX{{\rm B\kern-.05em{\sc i\kern-.025em b}\kern-.08em
    T\kern-.1667em\lower.7ex\hbox{E}\kern-.125emX}}
\begin{document}

\title{Byzantine-Robust Decentralized Coordination of LLM Agents}

% Double Blind Review
\author{\IEEEauthorblockN{Yongrae Jo}
    \IEEEauthorblockA{\textit{Department of Computer Science and Engineering} \\
        \textit{Pohang University of Science and Technology}\\
        Pohang, Republic of Korea \\
        memex@postech.ac.kr}
    \and
    \IEEEauthorblockN{Chanik Park}
    \IEEEauthorblockA{\textit{Department of Computer Science and Engineering} \\
        \textit{Pohang University of Science and Technology}\\
        Pohang, Republic of Korea \\
        cipark@postech.ac.kr}
}
\newcommand{\GMFaultThreshold}{$f \le \lfloor \frac{n-1}{2}\rfloor$}

\maketitle

\begin{abstract}
    Collaboration among multiple large language model (LLM) agents is a promising approach to overcome inherent limitations of single-agent systems, such as hallucinations and single points of failure. As LLM agents are increasingly deployed on open blockchain platforms, multi-agent systems capable of tolerating malicious (Byzantine) agents have become essential.
    Recent Byzantine-robust multi-agent systems typically rely on leader-driven coordination, which suffers from two major drawbacks. First, they are inherently vulnerable to targeted attacks against the leader. If consecutive leaders behave maliciously, the system repeatedly fails to achieve consensus, forcing new consensus rounds, which is particularly costly given the high latency of LLM invocations. Second, an underperforming proposal from the leader can be accepted as the final answer even when higher-quality alternatives are available, as existing methods finalize the leader's proposal once it receives a quorum of votes.
    To address these issues, we propose DecentLLMs, a novel decentralized consensus approach for multi-agent LLM systems, where worker agents generate answers concurrently and evaluator agents independently score and rank these answers to select the best available one. This decentralized architecture enables faster consensus despite the presence of Byzantine agents and consistently selects higher-quality answers through Byzantine-robust aggregation techniques.
    Experimental results demonstrate that DecentLLMs effectively tolerates Byzantine agents and significantly improves the quality of selected answers.
\end{abstract}


\section{Introduction}
LLMs \cite{gpt4o,qwen3,Claude,llama,google_gemini,deepseek} have demonstrated remarkable performance across a wide range of natural language processing tasks such as reasoning \cite{MindOfSociety}, summarization \cite{gao2023humanlikesummarizationevaluationchatgpt}, translation \cite{IsMadSilverBullet}, coding \cite{DACode}, smart contract auditing \cite{LLMSmartSec}, anomaly detection \cite{gai2023blockchainlargelanguagemodels}, medical diagnosis \cite{MaiDxo} and more.
One promising direction to enhance the capabilities of LLMs is through multi-agent systems \cite{MindOfSociety,ShouldWeBeGoingMAD,MoreAgentsIsAllYouNeed,BeyondSelfTalk,WhenOneLLMDrools,IsMadSilverBullet,ijcai2024p890}, in which multiple LLM-based agents collaborate to overcome the limitations of a single LLM, such as a single point of failure and hallucinations. In multi-agent systems, these agents jointly solve problems, improving answer quality through quorum-based voting or multi-round debates.

% LLMs have shown remarkable performance in various natural language processing tasks. LLM 의 enhance 시키는 방법으로 multi-agent system이 있는데, multi-agent system은 single LLM 의 limitations (e.g, single point of failure and hallucination) 을 극복하기 위해 여러 개의 에이전트가 협력하여, 가령 예를 들어, multi-round debate, majority voting 문제를 해결하는 시스템이다.
% Recently, byzantine-robust multi agent systems, such as BlockAgents \cite{BlockAgents} and Trusted MultiLLMN \cite{luo2025weightedbyzantinefaulttolerance}, have been proposed to tolerate byzantine failure of each LLM agent, 즉, LLM agent가 malicious 하게 behavior 해서 backdoor attack 이나 intentionally wrong fake answer 를 제공함으로써, answer quality 를 저하시킬 수 있는 것을 mitigate 하기 위한 것이다.

\begin{figure}[t]
    \centering
    \subfloat[\small Increasing Consensus Latency with Byzantine Leaders]{%
        \includegraphics[scale=0.40]{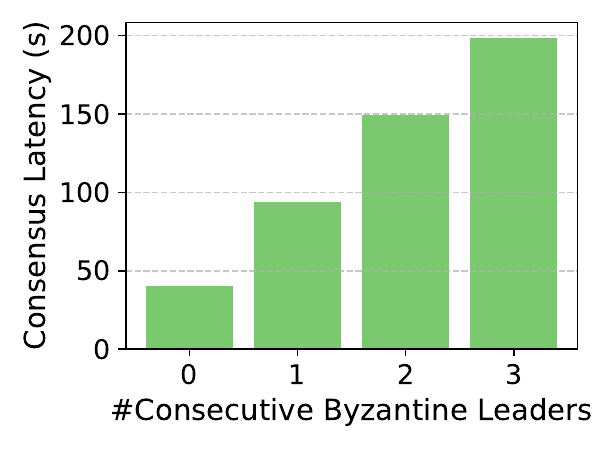}%
        \label{decentllms:consensus_latency}%
    }\hfill
    \subfloat[\small Accepting Underperforming Leader's Answer]{%
        \includegraphics[scale=0.40]{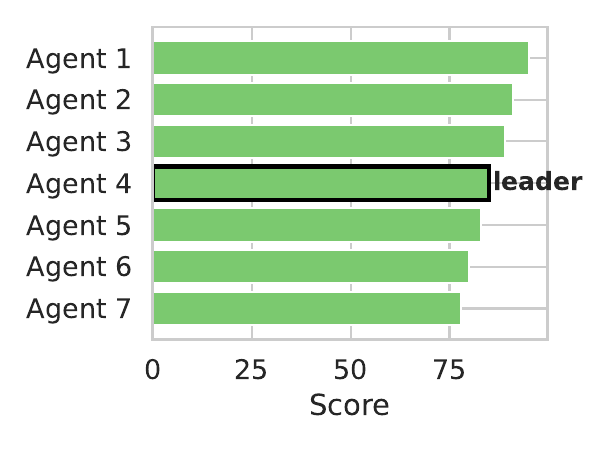}%
        \label{decentllms:underperforming_leader_answer}%
    }
    \caption{Motivating Examples: Leader-based consensus is vulnerable to Byzantine leaders and can finalize underperforming leader's answer due to quorum-based (e.g., majority) voting.}
    \label{fig:DecentLLMs:Motivation}
\end{figure}

Recently, open blockchain networks have seen the emergence of Byzantine-robust multi-agent systems \cite{BlockAgents, luo2025weightedbyzantinefaulttolerance} to tolerate Byzantine failures of individual LLM agents.
This is because, in a multi-agent system, a Byzantine agent can deliberately sabotage collaboration among agents, leading to poor-quality answers or increased latency during debate-style voting.
Existing Byzantine-robust multi-agent systems typically employ a leader-driven architecture, where a designated leader agent is responsible for coordinating the others to output a consensus answer. However, this architecture is inherently vulnerable to targeted attacks on the leader agent, which can significantly degrade the overall system performance.
For example, BlockAgents \cite{BlockAgents} designates a leader who first mines a block and then initiates a multi-round, debate-style voting process to obtain a majority of votes and reach consensus on the proposed block. If the leader is malicious and fails to secure a majority of votes within a preconfigured number of rounds, the protocol restarts the consensus process with a new leader.
Trusted MultiLLMN \cite{luo2025weightedbyzantinefaulttolerance} follows a HotStuff \cite{hotstuff}-like leader-rotation consensus mechanism that requires a two-thirds quorum of votes to succeed and designates a different leader in each round.
If the designated leader is Byzantine, the system restarts the consensus process in the next round with the new leader's proposal.

We identify two problems in such leader-based protocols.
First, consensus latency can increase significantly due to consecutive Byzantine leaders.
In these systems, consensus is attempted on a single leader's proposal each round. If the leader is Byzantine or submits a low-quality answer that fails to obtain a quorum, the round does not reach consensus and must be rerun with a new leader. In the worst case, multiple consecutive Byzantine leaders can dramatically increase latency, as illustrated in Fig. \ref{decentllms:consensus_latency}.
Second, the leader's proposal may not be the best answer available.
A quorum-based vote (e.g., majority or two-thirds) on a single leader's proposal can finalize an underperforming answer. For example, in Fig. \ref{decentllms:underperforming_leader_answer}, seven LLM agents generate responses with varying quality scores. Although better proposals exist, if Agent 4 is elected leader, its underperforming answer can be accepted through majority voting (i.e., votes from Agents 4-7), even though higher-quality alternatives are available.

\begin{figure}[t]
    \centering
    \subfloat[\small Single Answer Evaluation Consensus in a Round]{%
        \includegraphics[scale=0.325]{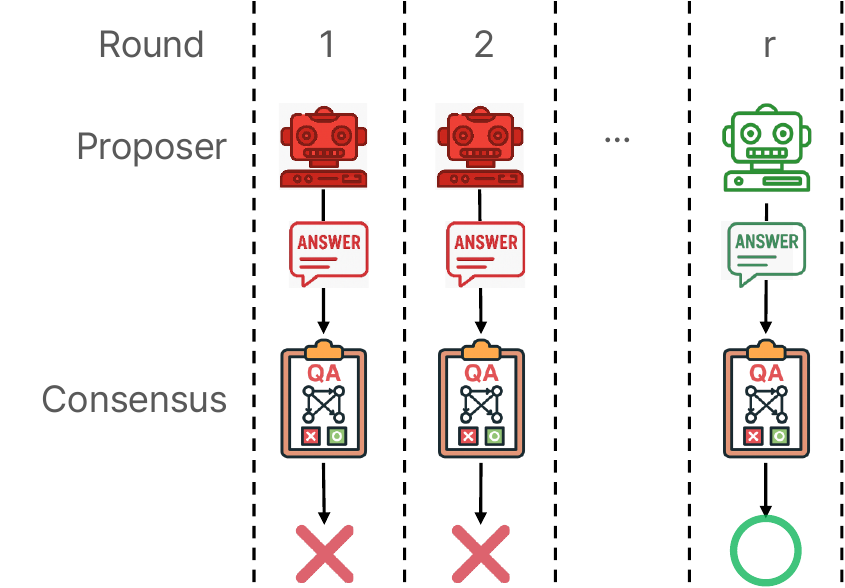}%
        \label{decentllms:fig:approach_1}%
    }\hfill
    \subfloat[\small Multiple Answers Evaluation Consensus in a Round]{%
        \includegraphics[scale=0.325]{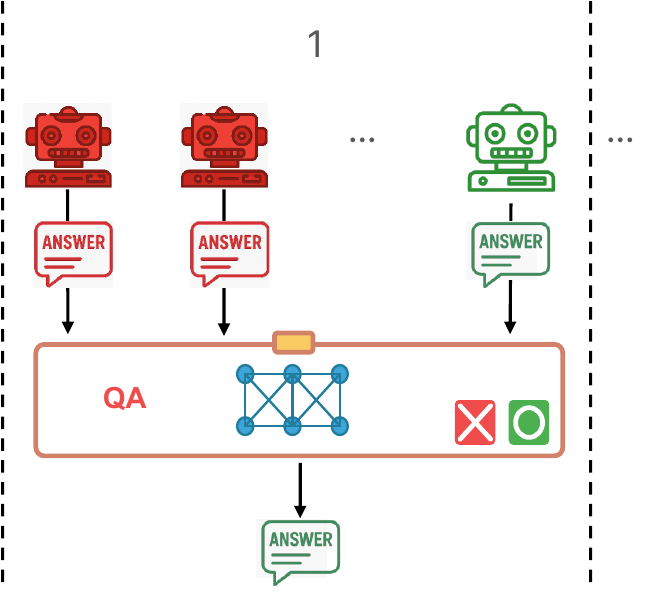}%
        \label{decentllms:fig:approach_2}%
    }
    \caption{Comparisons with leader-based and leader-less approaches in Byzantine-robust multi-agent systems.}
    \label{fig:DecentLLMs:Approaches}
\end{figure}

To address these limitations, we adopt a leaderless approach in which multiple answers are evaluated in parallel within a single round, and the highest-quality answer is selected. Unlike leader-based approaches that vote on one leader's answer per round, this method evaluates all participants' answers simultaneously, thereby mitigating the risk of poor-quality outputs and reducing consensus latency even in the presence of Byzantine agents. We illustrate the differences between leader-based and leaderless approaches in Fig. \ref{decentllms:fig:approach_1} and Fig. \ref{decentllms:fig:approach_2}.

To this end, we propose DecentLLMs, a decentralized multi-agent system that employs a leaderless consensus architecture: worker agents generate answers in parallel, and evaluator agents score them to select the highest-quality answer. This architecture enables all agents within each role to participate equally in the consensus process.
To achieve Byzantine-robust score aggregation by evaluator agents, DecentLLMs adopts the Geometric Median (GM) algorithm \cite{ByzantineMLPrimer}, which remains resilient to Byzantine attacks as long as an honest majority of evaluator agents exist.

\section{Background and Related Works}

\subsection{Multi LLM Agents Cooperation}
An LLM agent excels at generating fluent natural language, but it is also known to suffer from serious limitations such as hallucinations \cite{HallucinationSurvey}. For example, an agent may fabricate facts that were never in its training data, make logical inconsistencies, stray from the original question (loss of contextual relevance), or exhibit bias toward early tokens.
Multi-LLM agents have emerged as a promising way to address these issues \cite{MindOfSociety,ShouldWeBeGoingMAD,MoreAgentsIsAllYouNeed,BeyondSelfTalk,WhenOneLLMDrools,IsMadSilverBullet}. By running several LLM instances concurrently, the agents exchange answers, critique or debate one another, and iteratively refine their responses. Through peer feedback and quorum-based voting, they can produce higher-quality answers and reduce hallucinations.

\subsection{Malicious Byzantine Agents}
Existing multi-agent LLM frameworks do not take Byzantine behavior into account; most assume that errors are merely unintentional hallucinations. Yet, as AI agents are increasingly deployed in open peer-to-peer environments (e.g., decentralized GPU networks on public blockchains \cite{ionet,akashnetwork,bittensor}), Byzantine agents that can threaten the protocol are expected to proliferate. A Byzantine agent can inject inaccurate or nonsensical answers, disrupt cooperative protocols, and mislead end-users via backdoor attacks \cite{LLMSecuritySurvey2,LLMSecuritySurvey4}. Therefore, for multi-LLM agent systems operating in adversarial settings, Byzantine-robust mechanisms for detecting and mitigating such agents are essential.

\begin{figure}[t]
    \centering
    \subfloat[\small Multi Round Fixed Leader-driven Debate Style Consensus (e.g., \cite{BlockAgents})]{%
        \includegraphics[scale=0.6]{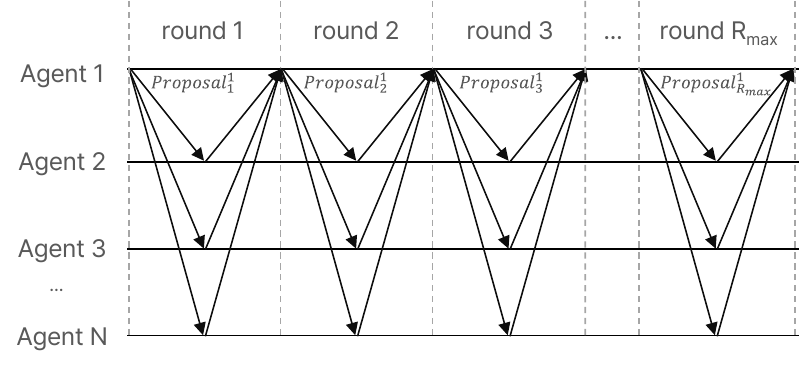}%
        \label{decentllms:fig:leader_driven_blockagents}%
    }\hfill
    \subfloat[\small Multi Round Rotating Leader-driven Quality Consensus (e.g., \cite{luo2025weightedbyzantinefaulttolerance})]{%
        \includegraphics[scale=0.6]{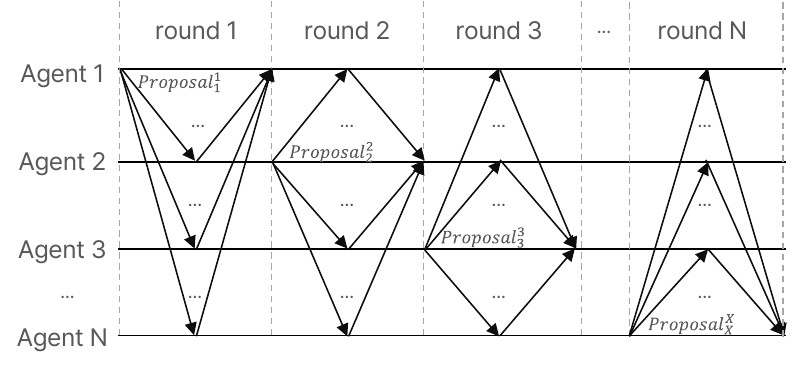}%
        \label{decentllms:fig:leader_driven_wbft}%
    }
    % \hfill
    % \subfloat[\small Multi Round Decentralized Consensus (our approach)]{%
    %     \includegraphics[scale=0.6]{figs/our_approach.pdf}%
    %     \label{decentllms:fig:our_approach}%
    % }
    \caption{Operations of Existing Leader-Based Consensus in Byzantine-Robust Multi-Agent Systems.}
    \label{fig:DecentLLMs:ExsistingApproaches}
\end{figure}

\begin{figure*}[t]
    \centering
    \includegraphics[scale=0.55]{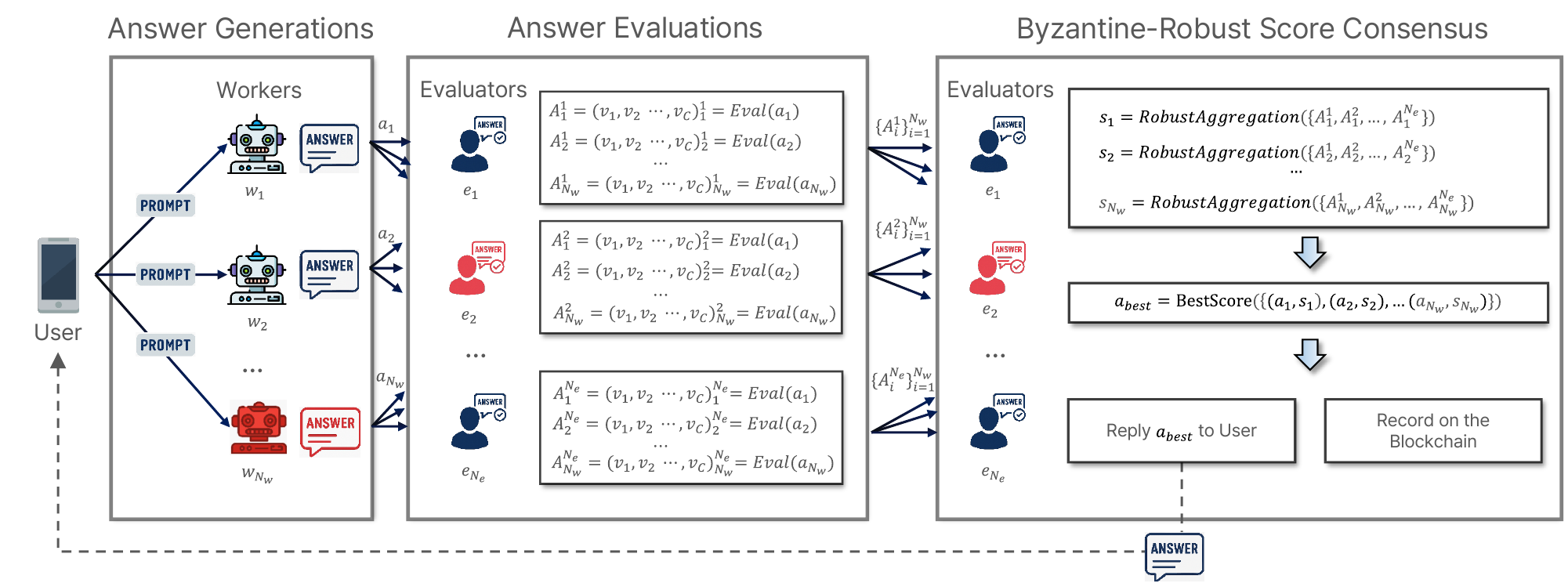}
    \caption{A protocol overview of DecentLLMs.}
    \label{fig:DecentLLMs:overview}
\end{figure*}

\subsection{Blockchain-based Byzantine-robust Multi-Agent Systems}
BlockAgents \cite{BlockAgents} and Trusted MultiLLMN \cite{luo2025weightedbyzantinefaulttolerance} are two recent systems that have been proposed to address the challenges of Byzantine agents in multi LLM agent systems.
BlockAgents is a framework that coordinates multiple LLM agents to solve a task collaboratively while adding a blockchain layer for tamper-evident, auditable records and incentive mechanisms. At startup, each agent is assigned one of two roles: workers, who propose answers, and evaluators (or miners), who judge those answers and record the results on-chain. When workers generate and submit their responses, evaluators attempt to build a block that contains their assessments and the corresponding rewards. The evaluator who first creates and broadcasts a valid block becomes the leader. The leader's block proposal then undergoes a multi-round, debate-style process that requires a majority vote by the other evaluators; once consensus is reached, the block is appended to the blockchain. We illustrate the multi-round, fixed-leader, debate-style consensus mechanism in Fig. \ref{decentllms:fig:leader_driven_blockagents}.

Similarly, Trusted MultiLLMN \cite{luo2025weightedbyzantinefaulttolerance} is an LLM-based multi-agent system that solves tasks collaboratively. The framework introduces Weighted Byzantine Fault Tolerance (WBFT), replacing the one-agent-one-vote rule with reputation-weighted voting. In each round, the protocol leader initiates a HotStuff-like two-phase prepare/commit process \cite{hotstuff}, during which followers compare the leader's answer with their own and cast weighted votes only if they deem it to be of better quality. When the cumulative weight in favor of the proposal exceeds two-thirds of the total honest stake, consensus is reached and the block containing the agreed answer is appended to the blockchain, ensuring tamper-evident and auditable records. We also depict the multi-round, rotating-leader, quality-based consensus mechanism in Fig. \ref{decentllms:fig:leader_driven_wbft}.

\section{DecentLLMs}
\subsection{Assumptions and Models}
Each agent possesses a public-private key pair that is used to sign and verify messages. Agents communicate over point-to-point links, assuming a synchronous network in which the maximum communication delay between agents is bounded and predetermined. We assume that standard cryptographic primitives are secure and have not been compromised.
Each agent in DecentLLMs is classified as either honest or Byzantine. Honest agents faithfully follow the DecentLLMs protocol.
Byzantine agents may behave arbitrarily, including the generation of extremely poor-quality answers (e.g., inserting advertisements), biased evaluations (e.g., assigning full/zero scores to Byzantine/honest agents, respectively), failing to respond, launching fork attacks (i.e., sending different values to different agents), or even colluding with other Byzantine agents to maximize their influence on the final output.
Note that even honest agents may produce relatively poor-quality answers owing to the inherent limitations (i.e., hallucinations) of LLMs; however, they are assumed not to generate overly anomalous answers beyond what would be expected under a known probability distribution. Users are assumed to be honest.

In DecentLLMs, there are two types of agents. A \emph{worker} agent $w$ in a worker group $W = \lbrace w_1, w_2, \ldots, w_{N_w} \rbrace$, where $N_w = |W|$, takes a user's prompt as input and generates answers by invoking an LLM. An \emph{evaluator} agent $e$ in an evaluator group $E = \lbrace e_1, e_2, \ldots, e_{N_e} \rbrace$, where $N_e = |E|$, evaluates the answers proposed by the worker agents and engages in a decision-making process to select one among them. It is assumed that both the worker group and the evaluator group have an honest majority. Formally, this implies $f_w < \lfloor \frac{N_{w}-1}{2}\rfloor$ and $f_e < \lfloor \frac{N_{e}-1}{2} \rfloor $, where $f_w$ and $f_e$ represent the maximum number of tolerable Byzantine agents within the worker and evaluator groups, respectively.
% We assume that the numerical evaluations provided by honest evaluators are independently and identically distributed (i.i.d.) according to a normal distribution with mean $\mu$ and standard deviation $\rho$.

% 우리는 honest 한 evaluators 들의 한 answer 에 대한 numerical 평과 결과들은 independent and identically distributed (iid) 를 따른다, 즉 honest evaluators 들의 평균은 $\mu$ 이고 편차는 $\rho$ 이다. 최대 평균 분산은 $\epsilon$ 이하라고 가정한다.

\subsection{Architecture}

We explain the design rationale behind the architecture of DecentLLMs. As previously mentioned, DecentLLMs does not rely on a leader agent; instead, multiple agents within each group participate equally in every phase of the protocol.

First, we adopt a \textit{parallel worker architecture}. In DecentLLMs, worker agents independently generate responses in parallel, based on the user's prompt, without communicating with each other. Generating multiple answers concurrently using several LLM instances for the same prompt is known to improve response quality by mitigating hallucinations \cite{MoreAgentsIsAllYouNeed}. This design supports horizontal scalability: by allocating additional GPU resources and increasing the number of worker agents, the system can produce more answers simultaneously.

Second, we employ a \textit{Byzantine-robust evaluation} process. To identify the highest-quality response among those generated by the worker agents, DecentLLMs utilizes evaluator agents that independently assess each answer. Their evaluations are then aggregated through a Byzantine-robust consensus mechanism to determine the best answer. This ensures that the best-performing answers are selected by scoring and ranking the worker agents' responses. Moreover, DecentLLMs leverages the geometric median (GM) algorithm \cite{ByzantineMLPrimer} to enhance robustness against Byzantine evaluators. As long as the majority of evaluator agents are honest, the geometric median of their scores reliably supports the selection of the best available answer.

% multi agents 에서는 multi-round debate in which, agent receives 피드백 to improve answer quality 방법 또한 존재한다. 하지만 이 방법은 합의 지연 시간을 높일 수가 있는데, DecentLLMs 이를 parallel worker architecture 를 통해 avoid 하고자 하지만, 그럼에도 불구하고, final  consensus decision 으로 결정된 답변이 threshold 보다 낮다면은, evaluators 들의 평가 결과를 다음 라운드의 worker agents 에게 전달하여, 그들의 개선된 답변을 유도하도록 만든다.
% Lastly, DecentLLMs 의 honest evaluators 들 끼리의 평가 결과 분산이 $\epsilon$ 이상이 되어서, minority 의 비잔틴 evaluation 쪽으로 byzantine 점수가 나올 수가 있는데, DecentLLMs 는 평가 결과 간의 분산이 $\epsilon$ 이상이 되면, evaluation 을 위한 새로운 라운드를 시작한다.

\subsection{Protocol}
We now describe the DecentLLMs protocol in detail. The protocol initiates when a user submits a prompt request to DecentLLMs. An overview of this consensus protocol is presented in Fig. \ref{fig:DecentLLMs:overview}. We explain the three phases, including \textit{answer generation}, \textit{answer evaluation}, and \textit{Byzantine-robust score consensus}. The protocol proceeds synchronously.

\subsubsection{Answer Generations}
In this phase, the user broadcasts a prompt (i.e., a transaction request) to the worker agents. Each worker agent $w_i$ independently invokes its own LLM instance to generate an answer $a_i$, subsequently broadcasting this answer to evaluator agents. A malicious worker agent may fail to broadcast an answer or may broadcast conflicting answers (i.e., a fork attack). To address such situations, DecentLLMs employs Byzantine reliable broadcast protocols over a synchronous network (e.g., \cite{SonicBFT}), ensuring that answers from honest worker agents are consistently broadcast to evaluator agents within bounded time. Note that because each worker agent may be equipped with different LLM instances, the latency of answer generation can vary, so the timeout for answer generation should be conservatively set.

\subsubsection{Answer Quality Evaluations}
Upon receiving the answers from workers, each evaluator agent $e_j$ evaluates each answer's quality using multiple criteria identified in \cite{HallucinationSurvey}. The criteria include: \textit{factual contradiction}, measuring the degree to which an answer contradicts established facts; \textit{factual fabrication}, measuring the degree to which an answer invents or fabricates non-existent facts; \textit{instruction inconsistency}, measuring the degree to which an answer deviates from or does not properly address the original prompt; \textit{context inconsistency}, measuring the extent to which an answer deviates from the original context of the question; and \textit{logical inconsistency}, measuring the degree of incorrect logical argumentation.
Each evaluation criterion is scored on a scale from 0 to 20, where scores closer to 20 indicate higher quality, scores closer to 0 indicate lower quality, and a score of 10 indicates neutrality.

As illustrated in Fig. \ref{fig:DecentLLMs:overview}, each evaluator agent $e_j$ invokes \textsf{Eval} function for each worker's answer, computing a quality score for each criterion. These scores are stored as a vector $A_i^j = (v_1, v_2, \ldots, v_C)_i^j$, where each $v$ represents the score for a criterion, and $C$ is the number of evaluation criteria. In our case, $C = 5$. If an evaluator agent is Byzantine, it may significantly distort the quality scores by assigning full scores to Byzantine agents’ answers or zero scores to honest workers' answers.

\subsubsection{Byzantine-Robust Score Consensus}
After evaluation, each evaluator agent $e_j$ broadcasts its evaluation vectors $\lbrace A_{i}^{j} \rbrace_{i=0}^{N_w}$ to other evaluators using Byzantine reliable broadcast. Consequently, each evaluator $e_j$ obtains a complete set of score vectors for each worker $w_i$'s answer $a_i$, denoted by $\lbrace A_i^1, A_i^2, \ldots, A_i^{N_e} \rbrace$.
Each evaluator then applies the geometric median (GM) algorithm \cite{ByzantineMLPrimer} to aggregate the evaluation vectors, some of which may be Byzantine. The GM finds the vector that minimizes the sum of Euclidean distances to all input vectors.
DecentLLMs adopts the GM algorithm as follows. Using the formula, each evaluator computes a robustly aggregated score $s_i$ for each worker $w_i$'s answer:

\begin{equation}
    s_{i} = \mathrm{GM}(A_{i}^{1}, A_{i}^{2}, \ldots, A_{i}^{N_{e}})  \in \mathop{\arg\min}_{z \in \mathbb{R}^C} \sum_{k=1}^n \| z - A_{i}^{k} \|
\end{equation}

% Formally, it is defined as follows:
% DecentLLMs adopts the GM algorithm as follows: each evaluator computes a robustly aggregated score $s_i$ for every worker $w_i$’s answer,

% \begin{equation}
%     \mathrm{GM}(x_1, \ldots, x_n) \in \mathop{\arg\min}_{z \in \mathbb{R}^d} \sum_{k=1}^n \| z - x_k \|
%     \label{eq:GM}
% \end{equation}

% Using this formula, each evaluator computes a robustly aggregated score $s_i$ for each worker $w_i$'s answer:

Note that we choose the GM algorithm because, compared to other Byzantine-robust aggregation algorithms such as Krum \cite{Krum} and Bulyan \cite{Bulyan}, GM offers higher Byzantine resilience. Specifically, GM tolerates up to \GMFaultThreshold Byzantine vectors, while Krum and Bulyan tolerate up to $f \le \lfloor \frac{n-2}{2}\rfloor$ and $f \le \lfloor \frac{n-3}{4}\rfloor$, respectively. Although the GM algorithm is known for its high computational complexity, in DecentLLMs this complexity is acceptable because the number of evaluation dimensions $C$ is small (e.g., five in our criteria) and the number of evaluation vectors $N_e$ is relatively manageable.
% Moreover, since evaluators perform GM aggregation over identical datasets (due to Byzantine reliable broadcast over a synchronous network), the aggregated score $s_i$ for each answer $a_i$ is computed deterministically and consistently across all honest evaluators.

After calculating the aggregated scores $s_i$ for all answers, the evaluators identify the best answer $a_{best}$ as the one with the highest robust score.
If there are multiple answers tied for the highest score, $a_{best}$ is selected as the one with the largest hash value among the results obtained by hashing the concatenation of each answer with the most recent block hash.
Each evaluator agent then replies to the user with this answer, ensuring the user receives consistent responses from a majority of evaluators. Additionally, each evaluator $e_j$ generates and records a transaction onto the blockchain, including the prompt request, worker answers, evaluation results, and their corresponding signatures. These records can later be utilized for reconfiguring the worker and evaluator groups.

\begin{figure*}[t]
    \centering
    \includegraphics[scale=0.75]{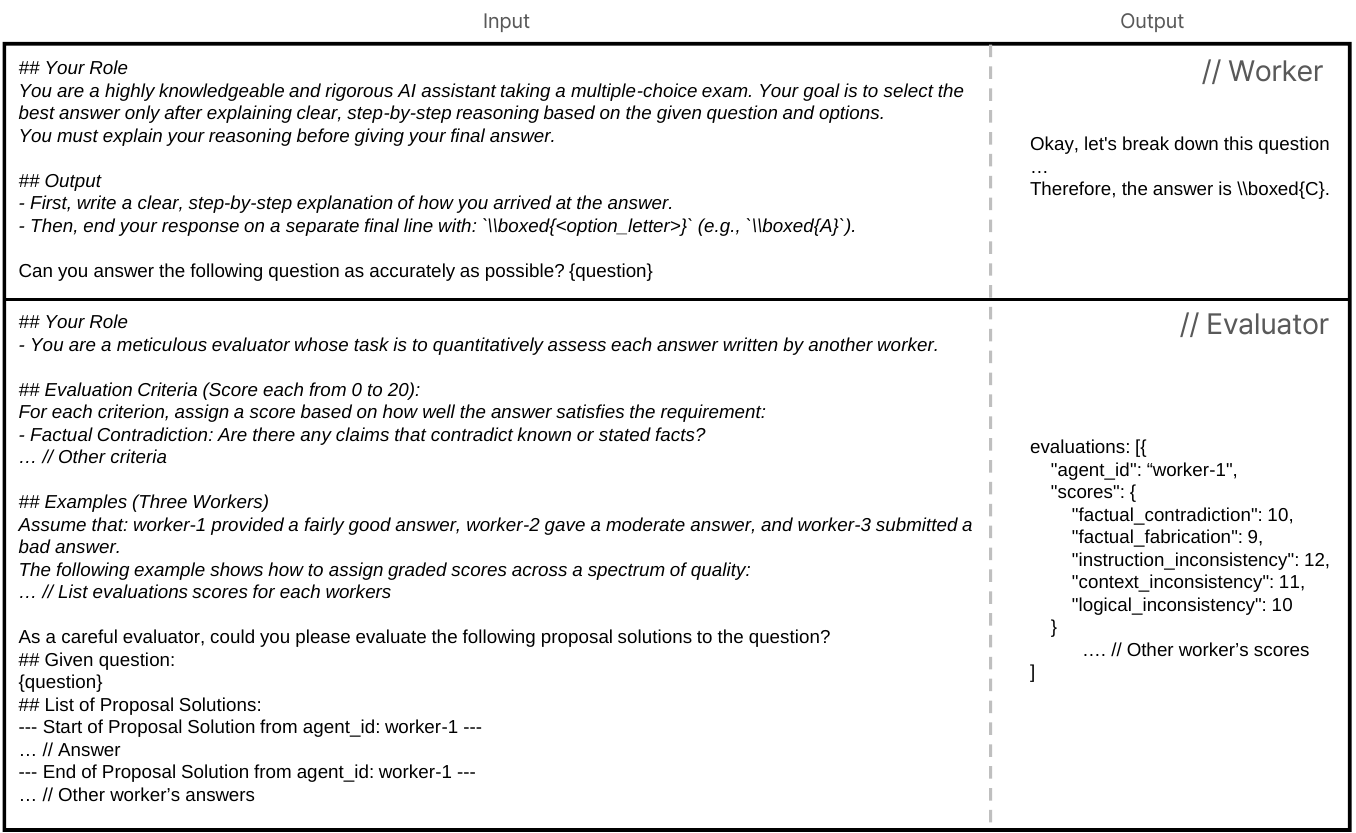}
    \caption{Prompts used in DecentLLMs. Role-playing prompting is applied throughout, with additional chain-of-thought prompting for workers and few-shot examples for evaluators. Evaluators produce JSON-structured outputs. Note that italicized text indicates system prompts, while normal text indicates user prompts in input boxes.}
    \label{decentllms:prompt_design}
\end{figure*}

\section{Implementation}
We outline the implementation of DecentLLMs. To effectively elicit high-quality outputs from LLMs, DecentLLMs employs several well-established prompting strategies, including role-play, chain-of-thought, and few-shot examples, as illustrated in Fig. \ref{decentllms:prompt_design}. To efficiently handle numerical outputs from evaluators, DecentLLMs adopts a JSON-format output for easy parsing and aggregation. The geometric median (GM) is computed using Weiszfeld's algorithm \cite{GMWeiszfeldAlgorithm}, with a maximum of 1000 iterations and a convergence tolerance of 1e-5. Communication between agents is implemented using gRPC.

\section{Evaluation}
In this section, we present the experimental setup and results evaluating the effectiveness of DecentLLMs. We experimentally address the following questions: 1) \emph{How much does DecentLLMs improve accuracy compared to existing quorum-based voting methods?} 2) \emph{How much does DecentLLMs reduce consensus latency compared to leader-based approaches in the presence of Byzantine agents?} 3) \emph{Does DecentLLMs assign appropriate scores to the generated answers?} 4) \emph{Does DecentLLMs exhibit sufficient Byzantine resilience?}

\subsection{Experimental Setups}
We prototype DecentLLMs in Python 3.11.11 \cite{Python:3.11.11}, supporting both worker and evaluator agents.
For worker agents, we deploy up to 9 agents using a variety of open-source LLMs, including Qwen3 (0.6B, 1.7B, 4B, 8B, 14B) and Gemma3 (1B, 4B, 12B, 27B).
Each worker runs on a dedicated local machine equipped with an AMD Ryzen Threadripper 3990X (2.9 GHz), 256 GB RAM, and either an NVIDIA RTX 3080 or 3080 Ti GPU.
For evaluator agents, we employ more capable models, including Claude Sonnet (4, 3.7), Gemini 2.5 Pro, GPT (4o, o3, o4-mini), and Qwen3(235B, 32B).
We randomly sample user prompts from the MMLU-Pro benchmark dataset \cite{MMLUPro}, which contains a challenging problems for the recent LLM models.

\paragraph*{Byzantine Agents Simulation}
To simulate Byzantine workers, we intentionally manipulate the outputs of normal LLMs by inserting advertisements into responses and randomly altering numerical values.
Byzantine evaluators are simulated to collude with Byzantine workers by assigning full scores (i.e., 100) to the responses from Byzantine workers and zero scores to those from honest workers, thereby increasing the likelihood that the Byzantine responses are selected as the final outputs.

% We configured Ollama with $temperature=1.0$, $top\_k=200$, $top\_p=0.94$, and $num\_ctx=3k$.

% \begin{figure}[t]
%     \centering
%     \includegraphics[scale=0.6]{figs/Baseline_Experiment_MMLUPro2.pdf}
%     \caption{Workers configurations with heterogeneous LLM Models. We also show their performance using MMLU-Pro \cite{MMLUPro}.}
%     \label{decentllms:evaluation_baseline_experiment_mmlupro}
% \end{figure}
% For evaluator agents, we used a structured output format in which each agent rates the five evaluation criteria on a scale from $-10$ to $10$, making their responses easy to parse and aggregate.

% \subsection{Byzantine Agents Simulation}

% 실험 안에 Byzantine Resillience 내용도 같이 들어가야 할 듯.
\subsection{Results}

\subsubsection{Accuracy Improvements over Quorum-based Voting Methods}
\begin{figure}[t]
    \centering
    \includegraphics[scale=0.8]{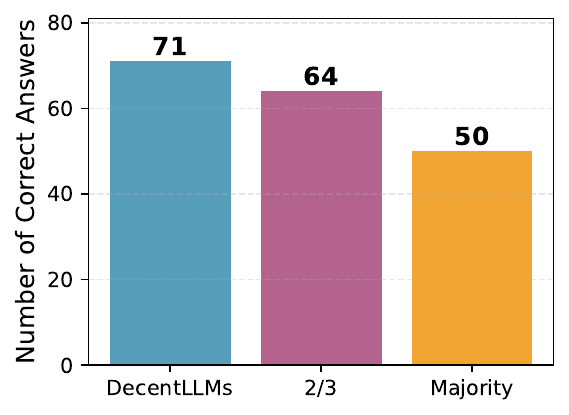}
    \caption{Accuracy comparisons of DecentLLMs and other quorum-based voting methods, including 2/3-quorum and majority quorum. We selected 100 random problems from MMLU-Pro \cite{MMLUPro} benchmark dataset.}
    \label{decentllms:evaluation_quorum_method_comparisons}
\end{figure}

We present an experimental comparison of DecentLLMs with quorum-based voting methods used in leader-based protocols, including 2/3-quorum (e.g., \cite{luo2025weightedbyzantinefaulttolerance}) and majority quorum (e.g., \cite{BlockAgents}), as shown in Fig. \ref{decentllms:evaluation_quorum_method_comparisons}.
For simplicity, in both 2/3-quorum voting and majority quorum voting, we assumed that the leader's answer proposal corresponds to the $\lceil \frac{(N_{w}+1)}{3} \rceil$-th and $\lceil \frac{(N_{w}+1)}{2} \rceil$-th ranked answers among all worker responses, respectively, since these positions would receive sufficient votes from agents with lower-scoring answers.
We used 100 random problems from the MMLU-Pro \cite{MMLUPro} benchmark dataset for evaluation.
As a result, DecentLLMs outperforms both quorum-based voting methods, achieving an accuracy of 71 — representing a 7\% improvement over 2/3-quorum (64) and a 21\% improvement over majority quorum (50), respectively.
This performance difference is mainly due to DecentLLMs' approach of evaluating and ranking all worker answers to select the highest-scoring response as the final output. In contrast, the other two methods select the leader's underperforming proposal as the final output as long as it receives sufficient quorum votes.

\subsubsection{Consensus Latency}
\begin{figure}[t]
    \centering
    \includegraphics[scale=0.8]{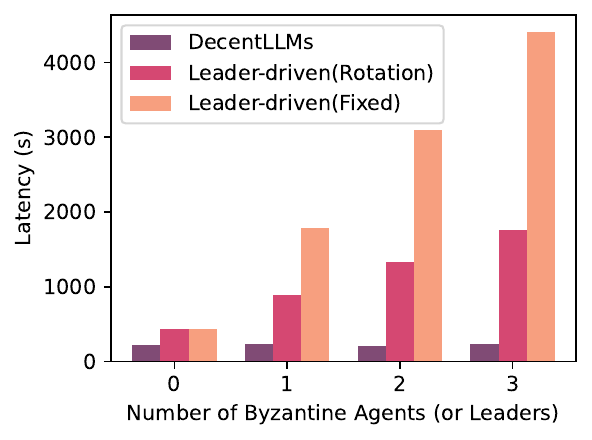}
    \caption{Consensus Latency of DecentLLMs and Leader-based Approaches.}
    \label{decentllms:evaluation_consensus_latency}
\end{figure}

We compare the consensus latency of DecentLLMs for a single user request with two leader-driven baselines, including fixed leader, debate-style consensus (e.g., \cite{BlockAgents}) and rotating leader-driven quality consensus (e.g., \cite{luo2025weightedbyzantinefaulttolerance}) in terms of the number of Byzantine agents. We simulate their communication patterns as shown in Fig. \ref{fig:DecentLLMs:Approaches} and DecentLLMs's consensus latency is measured as the time taken from evaluators all-to-all broadcasting their evaluation results.
For the leader-driven approaches we assume the worst case in which Byzantine leaders are selected in successive rounds. In the fixed-leader debate protocol, each leader run at most three consensus rounds.
As shown in Fig. \ref{decentllms:evaluation_consensus_latency}, DecentLLMs maintains a roughly constant latency of around 221 seconds, regardless of the number of Byzantine agents.
In contrast, both leader-driven approaches exhibit a nearly linear increase in latency as the number of Byzantine agents increases.
This difference occurs because DecentLLMs' decentralized consensus requires only a single round of communication among evaluators, even as the number of Byzantine agents grows.
By contrast, the two leader-driven approaches require the system to either (i) rotate to a new leader in the next round (the rotating-leader approach) or (ii) retry with the same leader for up to three rounds (the fixed-leader approach).

\begin{table*}[t]          % [t] : top of the page
    \centering              % 중앙 정렬
    %\caption{총합 기준 Robust 점수}   % (선택) 설명
    \label{tab:robust-scores}        % (선택) 참조용 라벨
    \begin{tabular}{lrrrrrrrrrrrr}
        \toprule
        \multicolumn{1}{r}{Worker ID} & $w_{0}$       & $w_{1}$       & $w_{2}$       & $w_{3}$       & $w_{4}$       & $w_{5}$       & $w_{6}$       & $w_{7}$       & $w_{8}$       & $w_{9}^{*}$     \\
        Evaluator ID                  &               &               &               &               &               &               &               &               &               &               & \\
        \midrule
        $e_{0}$ (Sonnet 4)            & 63            & 95            & 92            & 75            & 46            & 50            & 92            & 96            & 95            & 6               \\
        $e_{1}$ (Sonnet 3.7)          & 21            & 98            & 99            & 55            & 45            & 41            & 89            & 99            & 100           & 22              \\
        $e_{2}$ (Gemini 2.5)          & 33            & 100           & 75            & 45            & 40            & 35            & 99            & 100           & 85            & 0               \\
        $e_{3}$ (GPT-4o)              & 60            & 80            & 100           & 80            & 50            & 94            & 82            & 82            & 82            & 38              \\
        $e_{4}$ (GPT-o3)              & 52            & 83            & 66            & 57            & 44            & 84            & 84            & 83            & 66            & 35              \\
        $e_{5}$ (GPT-o4-mini)         & 49            & 53            & 60            & 19            & 42            & 27            & 66            & 66            & 66            & 7               \\
        $e_{6}$ (Qwen3-235B)          & 31            & 52            & 79            & 49            & 27            & 17            & 79            & 84            & 49            & 34              \\
        $e_{7}$ (Qwen3-32b)           & 75            & 100           & 80            & 100           & 50            & 64            & 100           & 100           & 100           & 39              \\
        $e_{8}^{*}$ (Byzantine)       & 0             & 0             & 0             & 0             & 0             & 0             & 0             & 0             & 0             & 100             \\
        \bottomrule
        \textbf{Robust Score}         & \textbf{45.1} & \textbf{83.8} & \textbf{80.4} & \textbf{55.6} & \textbf{43.5} & \textbf{42.0} & \textbf{84.5} & \textbf{88.2} & \textbf{83.0} & \textbf{28.5}   \\ \bottomrule
        \textbf{Correct Answer}       & \textbf{X}    & \textbf{O}    & \textbf{O}    & \textbf{O}    & \textbf{X}    & \textbf{X}    & \textbf{O}    & \textbf{O}    & \textbf{X}    & \textbf{X}      \\ \bottomrule
    \end{tabular}
    \caption{Evaluator's Scores on Worker Answers. The notations $w_{i}^{*}$ and $e_{j}^{*}$ denote a Byzantine worker and a Byzantine evaluator, respectively.}
\end{table*}

% \begin{figure}[t]
%     \centering
%     \includegraphics[scale=0.6]{figs/worker_score.pdf}
%     \caption{Worker Robust Scores.}
%     \label{decentllms:evaluation_worker_scores}
% \end{figure}

\subsubsection{Answer Quality Evaluation}
We demonstrate a snapshot that illustrates how DecentLLMs operates in a specific example execution, as shown in Fig. \ref{tab:robust-scores}, even in the presence of Byzantine agents. In the table, we show the received scores of each worker's (i.e., $w_{i}$) answers from each evaluator agent (i.e., $e_{j}$). The notations $w_{i}^{*}$ and $e_{j}^{*}$ denote Byzantine workers and evaluators, respectively. The last two rows show the robust score of each worker's answer, calculated by the GM algorithm, and whether each worker's answer is correct.
In the table, we observe that $w_{7}$ received the highest robust score of $88.2$, was selected as the final output, and provided the correct answer. The workers are ranked as follows: $w_{7}, w_{6}, w_{1}, w_{8}, w_{2}, w_{3}, w_{0}, w_{4}, w_{5}, w_{9}^{*}$.
The next two high-performing answers, $w_{6}$ and $w_{1}$, also provided correct answers. However, the the next ranked answer from $w_{8}$ was incorrect, indicating that the evaluators failed to reliably score the answers.
Additionally, although $w_{3}$ received a relatively low score, it still provided the correct answer, which further highlights the evaluators' reliability issues to assess answers.
Although DecentLLMs successfully selected the correct answer in this case, it also highlights the need to improve the reliability of the evaluators' scoring mechanism as an important direction for future work.

\subsubsection{Byzantine Resilience}
\begin{figure}[t]
    \centering
    \includegraphics[scale=0.8]{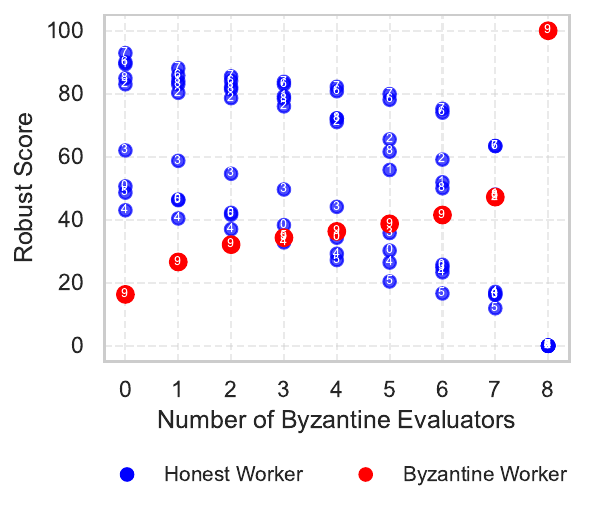}
    \caption{Number of Byzantine Evaluators. With eight honest evaluators, we incrementally add up to seven Byzantine evaluators. The DecentLLMs continues to select the answer from the honest workers until six Byzantine evaluators are present. However, when the number reaches seven—violating the GM algorithm's tolerance threshold (i.e., $f  < \lfloor \frac{15}{2} \rfloor $)—the DecentLLMs incorrectly selects the Byzantine worker's answer (i.e., from $w_{9}^{*}$)}.
    \label{decentllms:evaluation_byzantine_evaluator}
\end{figure}

Next, we evaluate the Byzantine resilience of DecentLLMs in the presence of Byzantine evaluators, as shown in Fig. \ref{decentllms:evaluation_byzantine_evaluator}.
In this experiment, we extend the previous example by gradually increasing the number of Byzantine evaluators up to seven to observe how the ranking based on robust scores changes under increasing Byzantine influence.
In the figure, honest workers are shown as blue circles and Byzantine workers as red circles. The number inside each circle denotes the corresponding worker's ID.
As a result, DecentLLMs, which employs the GM algorithm, successfully selects the correct answer up to the algorithm's theoretical fault tolerance threshold (i.e., 6). Beyond this point, however, the answer from the Byzantine worker (i.e., $w_{9}^{*}$) is selected.
Importantly, this experiment also implies that the quality of the workers' answers (and their corresponding evaluation scores) must be sufficiently high to fully leverage the Byzantine fault tolerance threshold of the GM algorithm.
As observed in the case with three to six Byzantine evaluators, some honest workers' answers received lower scores than those from the Byzantine worker, which could potentially prevent the correct answers from being selected in some cases.

% But, note that the 이 실험은 또한 , but fails to do so when the number of Byzantine evaluators reaches seven.

% Although the GM algorithm is theoretically tolerant to up to 7 Byzantine evaluators in this setup, our results show that DecentLLMs correctly selects an honest worker's answer only up to 5 Byzantine evaluators. When 6 or more Byzantine evaluators are introduced, the final output is the answer from a Byzantine worker ($w_{9}^{*}$).
% This discrepancy arises because even honest evaluators exhibit considerable variation in their scoring due to the inherent limitations of LLM-based evaluation. Such variance may result in incorrect evaluations, such as assigning high scores to low-quality responses, which weakens the practical effectiveness of the GM algorithm despite meeting the theoretical threshold.
% To address this, we believe that reducing disagreement among honest evaluators—for example, by iteratively minimizing scoring differences through multi-round debates—is necessary to fully leverage the resilience guarantees of the GM algorithm. We leave this direction as future work.
% In the meantime, we suggest mitigating the issue by removing underperforming evaluators from the committee based on reputation over blockchain networks, or by applying weighted evaluation schemes to reduce their influence.

\section{Discussion and Future Works}

\paragraph*{Leader-based and Leader-less Approach.}
Traditionally, state machine replication (SMR) in Byzantine consensus has been implemented using leader-based approaches (e.g., PBFT \cite{castro1999practical}). However, leader-based protocols often suffer from several limitations, including performance bottlenecks, the high complexity of leader replacement (i.e., view change protocol), and unfair transaction ordering. To address these issues, leader-less approaches (e.g., Narwhal \cite{narwhaltusk}) have been proposed, enabling all participants to engage equally in the consensus process.
DecentLLMs demonstrates the effective adoption of leader-less consensus architecture in LLM-based multi-agent systems, improving robustness against Byzantine agents and reducing consensus latency in worst-case scenarios compared to existing leader-based approaches \cite{BlockAgents, luo2025weightedbyzantinefaulttolerance}.

\paragraph*{Multi-Round Debates.}
In multi-LLM agent systems, multi-round debate \cite{MindOfSociety,ShouldWeBeGoingMAD,IsMadSilverBullet,Xiong_2023,MaiDxo} is a technique known for progressively improving the quality of agent responses and reducing the degree of disagreement by exchanging feedback over several rounds.
In DecentLLMs, multi-round debates could be applied in two ways. First, to enhance the quality of worker responses. This can be achieved by having workers exchange their answers to improve their quality, or by refining their answers based on feedback from evaluators.
Second, to reduce the variance among evaluator scores. High variance in scores, even among honest evaluators, can lead to vulnerability to Byzantine influence due to the mechanism of the Geometric Median (GM) algorithm. In our specific case, this effect was not observed because the answers from Byzantine workers contained clear signs of malicious content, prompting sufficiently low scores for effective classification. However, in scenarios with high score variance where Byzantine worker answers might receive high scores, multi-round debate could be employed.

\paragraph*{AI Blockchain Oracle.}
Blockchain oracles \cite{breidenbach2021chainlink, Supra, Kava,RedStone} are essential components that significantly improve the applicability of blockchain by providing external data feeds for smart contracts.
We consider utilizing DecentLLMs' LLM-powered multi-agent system as an AI-driven blockchain oracle, which represents a promising direction for future work.
A key challenge is to ensure reliable and consistent outputs from non-deterministic LLM models for smart contract execution.

% \paragraph*{Incentive Mechanisms.}
% The incentive mechanism typically provides rewards such as reputation and stakes to honest participants, while malicious participants are penalized (e.g., through slashing) to discourage harmful behavior.
% Previous approaches \cite{BlockAgents,luo2025weightedbyzantinefaulttolerance} use such incentive mechanisms to elect consensus committee members, increasing the likelihood that honest agents participate in consensus.
% Similarly, because DecentLLMs records histories of protocol operations on the blockchain for after-the-fact auditing, it can also support incentive mechanisms for reconfiguring both workers and evaluators to reduce Byzantine influence. The detailed mechanism remains as a future work.

% \paragraph*{Byzantine Robust Aggregation.}

% \paragraph{Synchronous Network Assumptions.}
% DecentLLMs 는 synchronous network 을 가정했는데, 그리고 이것은 모든 에이전트들은 synchronously 하게 각 스텝을 altogether 시작하는 것을 의미한다. 이러한 가정은 asynchronous network 으로의 확대를 가능하게 한다.

% \paragraph{AI Oracle.}

\section{Conclusion}
In this paper, we proposed DecentLLMs, a decentralized multi-agent system that adopts a leaderless consensus approach where worker agents generate answers in parallel and evaluator agents scores and ranking them to select best available answer using the Byzantine-robust aggregation algorithm. This design enables faster consensus and ensures that the highest-quality answer is selected, even in the presence of Byzantine agents.
We demonstrated the effectiveness and robustness of our approach through experimental evaluations.

% \section*{Acknowledgment}
% This work was supported by Institute of Information \& Communications Technology Planning \& Evaluation (IITP) grant funded by the Korea government(MSIT) (No. 2021-0-00484, Core Technologies for Hybrid P2P Network-based Blockchain Services)

\bibliographystyle{IEEEtran}  % IEEEtran 스타일을 사용
% \bibliography{latex-bib/sample-base,latex-bib/ai}
% \nocite{*}
\bibliography{sample-base}
% \bibliography{IEEEexample}

\end{document}